# A Performance Comparison of CUDA and OpenCL

Kamran Karimi    Neil G. Dickson    Firas Hamze

D-Wave Systems Inc.
100-4401 Still Creek Drive
Burnaby, British Columbia
Canada, V5C 6G9
{kkarimi, ndickson, fhamze}@dwavesys.com

**Abstract**
CUDA and OpenCL offer two different interfaces for programming GPUs. OpenCL is an open standard that can be used to program CPUs, GPUs, and other devices from different vendors, while CUDA is specific to NVIDIA GPUs. Although OpenCL promises a portable language for GPU programming, its generality may entail a performance penalty. In this paper, we compare the performance of CUDA and OpenCL using complex, near-identical kernels. We show that when using NVIDIA compiler tools, converting a CUDA kernel to an OpenCL kernel involves minimal modifications. Making such a kernel compile with ATI's build tools involves more modifications. Our performance tests measure and compare data transfer times to and from the GPU, kernel execution times, and end-to-end application execution times for both CUDA and OpenCL.

## 1. Introduction

Graphics Processing Units (GPUs) have become important in providing processing power for high performance computing applications. CUDA [7] and Open Computing Language (OpenCL) [11] are two interfaces for GPU computing, both presenting similar features but through different programming interfaces. CUDA is a proprietary API and set of language extensions that works only on NVIDIA's GPUs. OpenCL, by the Khronos Group, is an open standard for parallel programming using Central Processing Units (CPUs), GPUs, Digital Signal Processors (DSPs), and other types of processors.

CUDA can be used in two different ways, (1) via the runtime API, which provides a C-like set of routines and extensions, and (2), via the driver API, which provides lower level control over the hardware but requires more code and programming effort. Both OpenCL and CUDA call a piece of code that runs on the GPU a *kernel*. There are differences in what each language accepts as a legal kernel, in our case making it necessary to change the kernel sources, as explained in section 2.

Setting up the GPU for kernel execution differs substantially between CUDA and OpenCL. Their APIs for context creation and data copying are different, and different conventions are followed for mapping the kernel onto the GPU's processing elements. These differences could affect the length of time needed to code and debug a GPU application, but here we mainly focus on runtime performance differences.

OpenCL promises a portable language for GPU programming, capable of targeting very dissimilar parallel processing devices. Unlike a CUDA kernel, an OpenCL kernel can be compiled at runtime, which would add to an OpenCL's running time. On the other hand, this just-in-time compile may allow the compiler to generate code that makes better use of the target GPU. CUDA, on the other hand, is developed by the same company that develops the hardware on which it executes, so one may expect it to better match the computing characteristics of the GPU, offering more access to features and better performance. Considering these factors, it is of interest to compare OpenCL's performance to that of CUDA in a real-world application.

In this paper we use a computationally-intensive scientific application to provide a performance comparison of CUDA and OpenCL on an NVIDIA GPU. To better understand the performance implications of using each of these programming interfaces, we measure data transfer times to and from the GPU, kernel execution times, and end-to-end application running times. Since in our case the OpenCL and CUDA kernels are very similar, and the rest of the application is identical, any difference in performance can be attributed to the efficiency of the corresponding programming framework.

Not much formal work has been done on systematic comparison of CUDA and OpenCL. An exception is [6], where CUDA and OpenCL are found to have similar performance. A benchmark suite that contains both CUDA and OpenCL programs is explained in [2]. A performance study for ATI GPUs, comparing the performance of OpenCL with ATI's Stream computing system [10] is outside the scope of this paper.

The rest of the paper is organized as follows. Section 2 presents our test application, as well as the OpenCL and CUDA kernels. Section 3 explains the performance tests we performed and analyzes the results. Section 4 concludes the paper.

**2. The application**

The application used in this paper, called Adiabatic QUantum ALgorthms (AQUA) [12], is a Monte Carlo simulation [1] of a quantum spin system written in C++. We approximate the quantum spin configuration with a classical Ising spin system [4]. The classical approximation consists of ferromagnetically-coupled copies of the quantum system. Each copy is coupled to exactly two other copies, forming a ring of the copies. This approximation process is called the Suzuki-Trotter decomposition [3]. Here we call the result a *layered system*. In this paper we simulate quantum systems ranging in size from 8 qubits (quantum bits) to 128 qubits, while the number of layers used to approximate the quantum system with a classical one is set to be 128 for all problem sizes. The number of variables in this layered system is the number of qubits in each layer, multiplied by the number of layers. During a Monte Carlo *sweep*, every variable in a layer is probabilistically flipped, so each sweep requires examining all the variables and updating the ones that get flipped.

We simulate each layered system at different points during an adiabatic quantum evolution [9]. At each point, a complete layered system is simulated, so the total number of variables processed by the application is the number of variables in each layered system, multiplied by the number of points used to simulate the adiabatic evolution, as in Table 1.

| Qubits | Layers | Simulation Points | Classical Spin Variables |
|---|---|---|---|
| 8 | 128 | 27 | 27,648 |
| 16 | 128 | 34 | 69,632 |
| 32 | 128 | 37 | 151,552 |
| 48 | 128 | 57 | 350,208 |
| 72 | 128 | 71 | 654,336 |
| 96 | 128 | 111 | 1,363,968 |
| 128 | 128 | 129 | 2,113,536 |

Table 1. Quantum system sizes and their corresponding classical sizes

The mapping of data structures to GPU and CPU threads in AQUA is presented in detail in [5], where a CUDA implementation of the algorithm is explained. For this paper we optimized the kernel's memory access patterns. We then ported the CUDA kernel to OpenCL, a process which, with NVIDIA development tools, required minimal code changes in the kernel itself, as explained below. Other related code, for example to detect and setup the GPU or to copy data to and from the GPU, needed to be re-written for OpenCL.

We assign each multi-processor in the GPU to sweep a layered system. For an 8-qubit system, for example, 27 layered systems are to be swept because we have 27 simulation points. We thus have 27 work groups (in OpenCL language) or thread blocks (in CUDA language).

Table 2 shows the changes we had to make to the CUDA kernel code in order for it to compile and run under OpenCL with NVIDIA tools. Note that the last change listed is because OpenCL prevents the use of the address of an array element to index into the array.

| Change | CUDA kernel | NVIDIA OpenCL kernel |
|---|---|---|
| Type qualifiers | Use __shared__, etc. | Use __local, etc. |
| GPU thread indexing | Use threadIdx, etc. | Use get_local_id(), etc. |
| Thread synchronizing | Use __syncthreads() | Use barrier() |
| Array referencing | float *a2 = &a1[index1]; // then use: a2[index2]; | index = index1 + index2 // then use: a1[index] |

Table 2. Changes necessary to make a CUDA kernel compile under NVIDIA's OpenCL

No other changes were necessary to make the compute kernel compile and run under OpenCL. The Mersenne-Twister [8] random number generator's kernel, called by the compute kernel as explained in [5], required similar changes to compile and run under OpenCL.

We also tried porting the code to ATI GPUs. Doing so involved many more changes to the kernel (in addition to the ones mentioned in Table 2), primarily due to the lack of global variable declarations in ATI's OpenCL, as mentioned in Table 3 below.

| Change | CUDA kernel | ATI OpenCL kernel |
|---|---|---|
| Memory allocation | __global float mt[SIZE] // use mt in kernel k() | __kernel k(__global float *mt) // pass mt as an argument |
| Intrinsic functions | __int_as_float() | as_float() |

Table 3. Changes necessary to make a CUDA kernel compile under ATI's OpenCL

With ATI's OpenCL development tools one cannot allocate memory statically. The memory must be allocated prior to calling the kernel and a pointer must be used to access it. Figure 1 shows the code to initialize Mersenne-Twister's data structures for NVIDIA's OpenCL tools.

```
__global unsigned int mt[MAX_RAND_CHAINS][NN][MAX_RAND_THREADS];
__global int mti[MAX_RAND_CHAINS][MAX_RAND_THREADS];

__kernel void ocl_init_rand(int seed) {

  mt[chain][0][thread]= seed + chain * MAX_RAND_THREADS * NN + thread;

  for (mti[chain][thread]=1; mti[chain][thread]<NN; mti[chain][thread]++) {
    mt[chain][mti[chain][thread]][thread] =
        (1812433253UL * (mt[chain][mti[chain][thread]-1][thread] ^
        (mt[chain][mti[chain][thread]-1][thread] >> 30)) + mti[chain][thread]);
  }
}
```

Figure 1. Mersenne-Twister initialization code for NVIDIA's OpenCL compiler

Figure 2 shows the code of Figure 1, changed to accept pointers to dynamically allocated memory. Passing the arrays as in mt[][NN][MAX_AND_THREADS] works under NVIDIA's tools but not under ATI's tools. As a result, the index calculation operations of

Figure 2 are needed to map the one-dimensional allocated arrays to three-dimensional arrays required by Mersenne-Twister. The code in Figure 2 compiles and runs under both NVIDIA and ATI tools.

```
__kernel void ocl_init_rand(int seed, __global unsigned int *mt,  __global int *mti) {
  int chain = get_global_id(0);  int thread = get_global_id(1);
  int base =  chain * MAX_RAND_THREADS * NN + thread;

  mt[base] = seed + base;

  int index = chain * MAX_RAND_THREADS + thread;
  for (mti[index]=1; mti[index]<NN; mti[index]++) {
    int index2 = base + mti[index] * MAX_RAND_THREADS;
    int index3 = base + (mti[index] - 1) * MAX_RAND_THREADS;
    mt[index2] = (1812433253UL * (mt[index3] ^ (mt[index3] >> 30)) + mti[index]);
  }
}
```

Figure 2. Mersenne-Twister initialization code for ATI and NVIDIA OpenCL compilers

Note that even though we were able to achieve source-level compatibility between ATI and NVIDIA, the resulting executables were not compatible with the other vendor's hardware, so currently achieving runtime compatibility is not possible.

To reduce the effects of coding patterns on performance tests, for the rest of the paper we use very similar CUDA and OpenCL kernels compiled with NVIDIA's development tools, as in Figure 1. The kernels contain a mix of integer, floating point, and logical operations, acting on different data structures. This complexity sets them apart from some other GPU applications, where the kernel is used for simpler operations such as adding or multiplying matrix elements.

## 3. Performance tests

We tested CUDA and OpenCL versions of our application on an NVIDIA GeForce GTX-260. Both CUDA and OpenCL development tools were at version 2.3. In [5], we were concerned with maintaining the responsiveness of the computer, and purposefully reduced the GPU's load to make sure the computer remains usable while the application is running. For the experiments in this paper, we aim for maximum performance, so we reduced the CPU code execution, as well as data copy portions of the run to a minimum and increased the GPU's load to the maximum. As a result, the computer's user interface was very sluggish during these tests. No interaction with the computer was attempted during the actual data-gathering runs to make sure the GPU's computing power remained dedicated to the AQUA application.

The application goes through the following steps during its run: (1) Setup the GPU (includes GPU detection, compiling the kernel for OpenCL, etc.) (2) Read the input, (3) copy data to the GPU, (4) Run the kernel on the GPU, (5) copy data back to the host, (6) process the returned data using the CPU and output the results.

Table 4 reports the total amount of time needed to copy data to and from the GPU and run the kernel (the sum of the time needed to perform steps 3, 4, and 5) as the GPU Operations Time. Both kernels performed 20,000 sweeps of the variables in each layered system. The End-To-End time in Table 4 shows the amount of time needed to run the whole application from the beginning to end, corresponding to time spent for steps 1 through 6. We solved each problem 10 times with both CUDA and OpenCL to get repeatable average times.

| Qubits | GPU Operations Time | | | | End-To-End Running Time | | | |
|---|---|---|---|---|---|---|---|---|
| | CUDA | | OpenCL | | CUDA | | OpenCL | |
| | avg | stdev | avg | stdev | avg | stdev | avg | stdev |
| 8 | 1.97 | 0.030 | 2.24 | 0.006 | 2.94 | 0.007 | 4.28 | 0.164 |
| 16 | 3.87 | 0.006 | 4.75 | 0.012 | 5.39 | 0.008 | 7.45 | 0.023 |
| 32 | 7.71 | 0.007 | 9.05 | 0.012 | 10.16 | 0.009 | 12.84 | 0.006 |
| 48 | 13.75 | 0.015 | 19.89 | 0.010 | 17.75 | 0.013 | 26.69 | 0.016 |
| 72 | 26.04 | 0.034 | 42.32 | 0.085 | 32.77 | 0.025 | 54.85 | 0.103 |
| 96 | 61.32 | 0.065 | 72.29 | 0.062 | 76.24 | 0.033 | 92.97 | 0.064 |
| 128 | 101.07 | 0.523 | 113.95 | 0.758 | 123.54 | 1.091 | 142.92 | 1.080 |

Table 4. GPU and application running times in seconds

To better understand the efficiency of CUDA and OpenCL in data transfer and kernel operations, Table 5 breaks down the GPU Operations Time into the Kernel Running Time (step 4), and the Data Transfer Time to and from the graphics device (steps 3 and 5; performed once for each problem).

| Qubits | Kernel Running Time | | | | Data Transfer Time | | | |
|---|---|---|---|---|---|---|---|---|
| | CUDA | | OpenCL | | CUDA | | OpenCL | |
| | avg | stdev | avg | stdev | avg | stdev | avg | stdev |
| 8 | 1.96 | 0.027 | 2.23 | 0.004 | 0.009 | 0.007 | 0.011 | 0.007 |
| 16 | 3.85 | 0.006 | 4.73 | 0.013 | 0.015 | 0.001 | 0.023 | 0.008 |
| 32 | 7.65 | 0.007 | 9.01 | 0.012 | 0.025 | 0.010 | 0.039 | 0.010 |
| 48 | 13.68 | 0.015 | 19.80 | 0.007 | 0.061 | 0.010 | 0.086 | 0.008 |
| 72 | 25.94 | 0.036 | 42.17 | 0.085 | 0.106 | 0.006 | 0.146 | 0.010 |
| 96 | 61.10 | 0.065 | 71.99 | 0.055 | 0.215 | 0.009 | 0.294 | 0.011 |
| 128 | 100.76 | 0.527 | 113.54 | 0.761 | 0.306 | 0.010 | 0.417 | 0.007 |

Table 5. Kernel execution and GPU data transfer times in seconds.

Table 6 shows the amount of data transferred between the GPU and the host. The same amount of data is copied from the host to the GPU (step 3), and from the GPU back to the host (step 5), so each of the steps 3 and 5 transfers half of the amount shown in Table 6.

| Qubits | Data Transferred |
|---|---|
| **8** | 649.05 |
| **16** | 1,633.32 |
| **32** | 3,553.44 |
| **48** | 8,210.22 |
| **72** | 15,338.77 |
| **96** | 33,124.49 |
| **128** | 49,541.04 |

Table 6. Amount of data transferred between the GPU and the host in KB.

To compare the data transfer times of CUDA and OpenCL, Figure 3 shows the transfer time for OpenCL divided by the transfer time for CUDA for each problem size. As can be seen, OpenCL's data transfer overhead does not change significantly for different problem sizes.

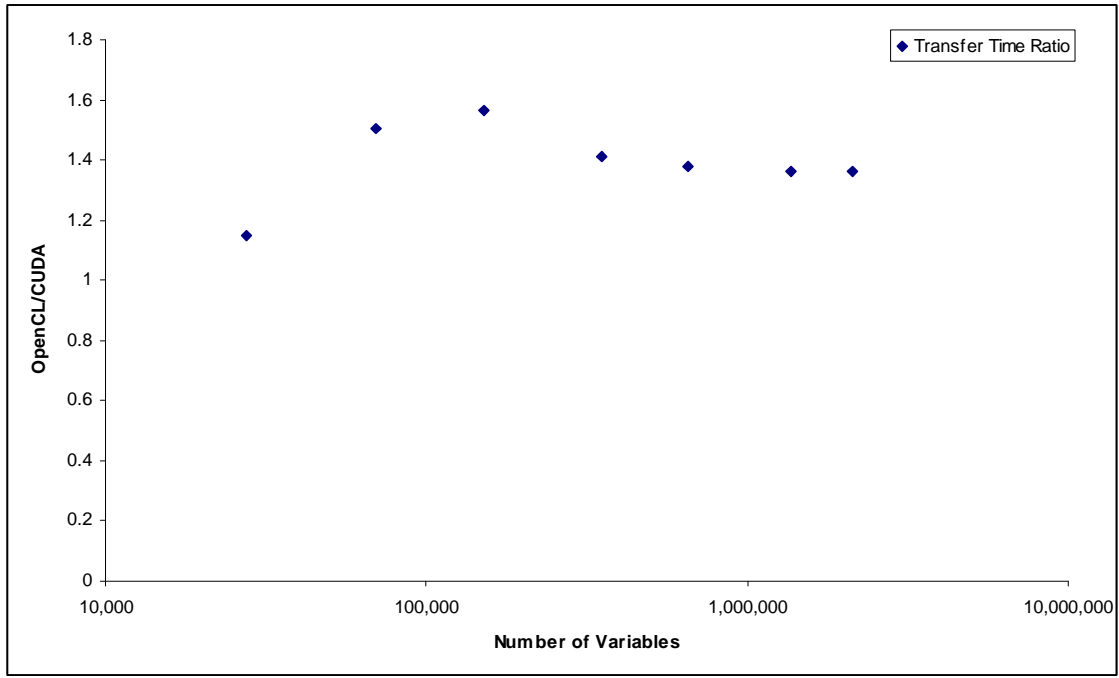

Figure 3. OpenCL/CUDA data transfer time ratio

Figure 4 displays the number of variables processed per second by the two kernels, as a function of the number of variables in the problem (Processed Variables / Kernel Running Time).

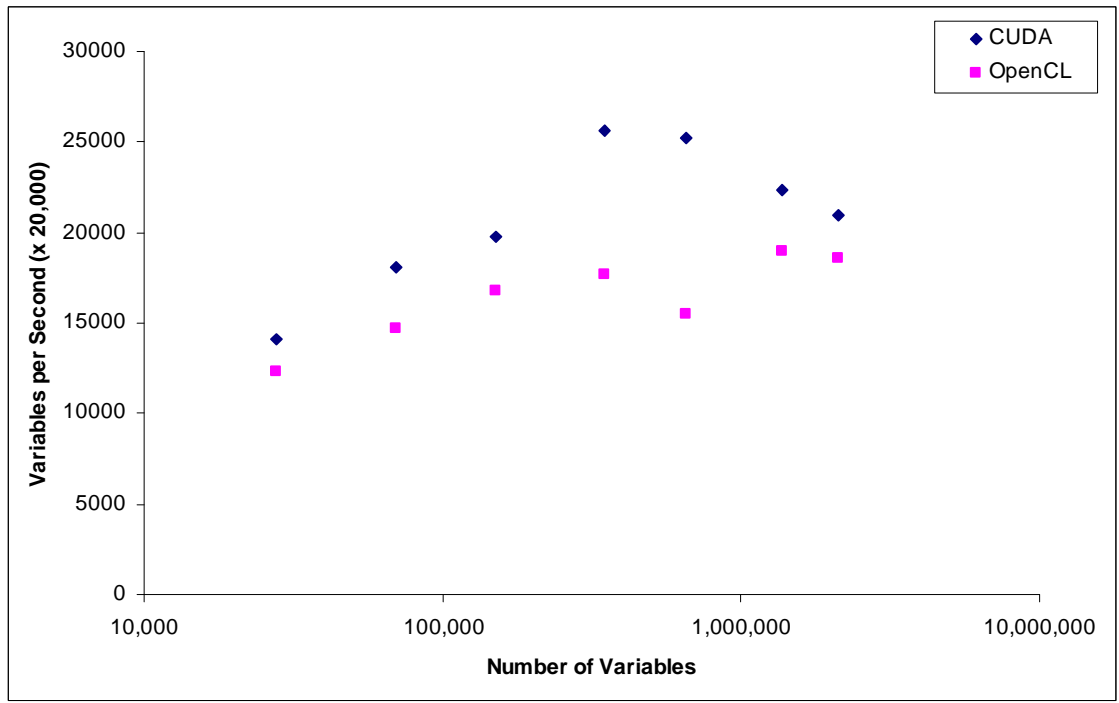

Figure 4. Processing speed for different problem sizes

From Figure 4 one can see that for each problem size, the CUDA version of the application processes more variables per seconds than the OpenCL version.

Figure 5 shows the relative time difference (i.e. (OpenCL's time – CUDA's time)/(CUDA's time)) for different problem sizes.. Data obtained from both the kernel execution time and the end-to-end running times are shown.

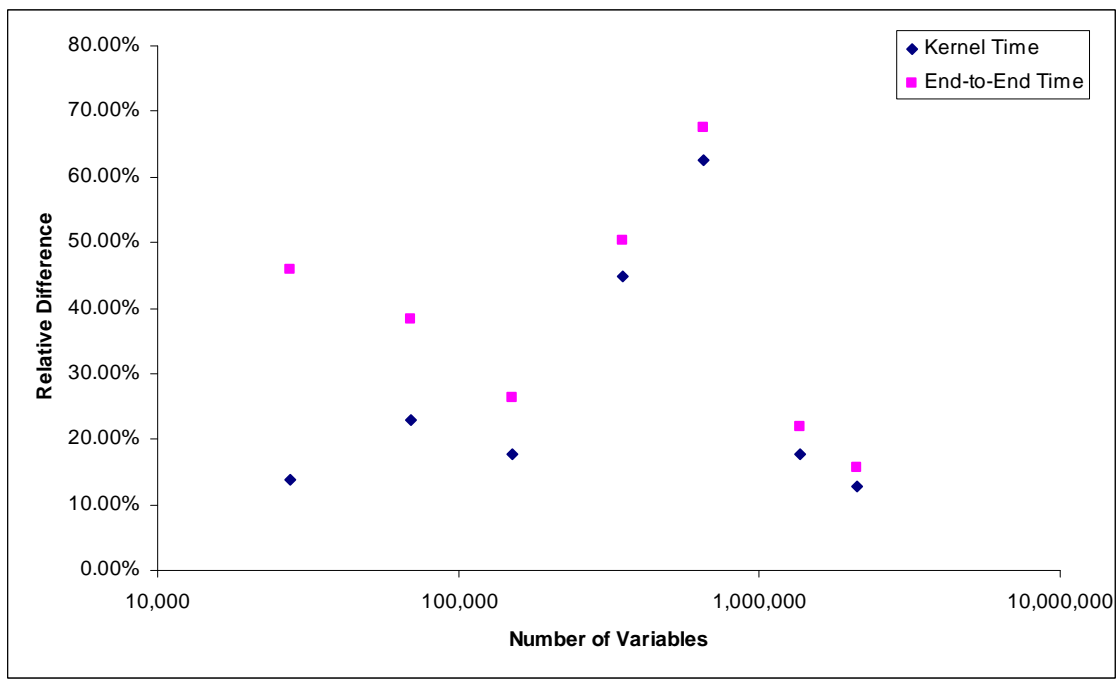

Figure 5. Relative difference in running time between CUDA and OpenCL

The changing performance for different problem sizes are due to differences in data structure sizes and their placement in GPU memory. GPU performance is very dependent on these issues. However, these effects are specific to the algorithm used, so here we focus on the performance difference between CUDA and OpenCL. For all problem sizes, both the kernel and the end-to-end times show considerable difference in favor of CUDA. The OpenCL kernel's performance is between about 13% and 63% slower, and the end-to-end time is between about 16% and 67% slower. As expected, the kernel and end-to-end running times approach each other in value with bigger problem sizes, because the kernel time's contribution to the total running time increases.

## 4. Concluding remarks

In this paper we used a specific real-world application to compare the performance of CUDA with NVIDIA's implementation of OpenCL. Both programming interfaces have similar functionality and porting the kernel code from one to the other needs minimal changes when using NVIDIA's development tools. Porting the rest of the GPU-related sources, including GPU setup and data transfer code, involves writing new code.

In our tests, CUDA performed better when transferring data to and from the GPU. We did not see any considerable change in OpenCL's relative data transfer performance as more data were transferred. CUDA's kernel execution was also consistently faster than OpenCL's, despite the two implementations running nearly identical code.

CUDA seems to be a better choice for applications where achieving as high a performance as possible is important. Otherwise the choice between CUDA and OpenCL can be made by considering factors such as prior familiarity with either system, or available development tools for the target GPU hardware.

**Acknowledgements**
We would like to thank Geordie Rose for supporting this project. We are also grateful to Corinna Klausing, David Lawson, and Tommy Sundgaard for their help.